\documentstyle[preprint,aps,eqsecnum]{revtex}

\begin{document}
\draft
\tighten

\title{QUANTUM FLUCTUATIONS OF SCALAR FIELD\\
IN CONICAL SPACE}
\author{Chung-I Kuo\\
Department of Physics\\
Soochow University\\
Taipei, Taiwan, Republic of China}
\vskip 1 true cm
\maketitle

\begin{abstract}
We consider vacuum polarization effect of a 
conformally coupled massless scalar field in the 
background produced by an idealized straight 
cosmic string. Using previous criterion we 
show the calculation of back reaction of the 
field to the metric in the context of
semiclassical gravity theory is not valid 
in some regions due to large quantum 
fluctuations in the conical space. 
\end{abstract}
\vskip 1 true cm
\vfill
\eject

\section{Introduction}

What we are going to consider is a scalar field
in the neighborhood of a straight, infinitely long
static cosmic string, taken as an example to see how 
large the deviation from the results expected 
from the semiclassical gravity theory will be. 

A cosmic string is one of several possible 
forms of topological defects formed during 
the phase transition in the early universe. 
There have been considerable interests in 
those objects especially due to the possibility 
that strings can serve as the seeds for galaxy 
formation(see \cite{Vilenkin85,Vilenkin94} for a full 
review).

However, a cosmic string is interesting in
itself in that the spacetime outside a
static infinite straight string is locally 
flat but globally conical with a deficit 
angle related to the linear mean density of the
string \cite{Vilenkin87}. Due to the boundary
(Casimir) effect \cite{Casimir48}, even the stress 
tensor of a free scalar field will not vanish in 
this spacetime.
Helliwell and Konkowski \cite{Helliwell86}
(see also \cite{Linet87,Dowker87,Dowker90,Smith90}) 
first calculated the effect of vacuum fluctuations of
a conformal scalar field outside a straight string.
The calculated stress tensor is traceless, falls 
off as the fourth power of the distance from the 
string, and is proportional to the linear mean 
density of the string in the limit of small linear 
mean density. Most important of all, the energy
density of the conformal scalar field is negative,
which is another example of the existence of negative
energy density in quantum field theory 
\cite{Epstein65,Kuo97}. Hiscock \cite{Hiscock87}
then calculated the stress tensor 
due to Casimir effect in the conical space in the 
vicinity of an infinitely long straight cosmic 
string, then used the semiclassical 
gravity theory to determine the back-reaction to 
the background metric itself. The calculations are 
very similar to the calculations of quantum 
fields in the wedge formed by two perfectly
conducting plates \cite{Deutsch79}.
He found out there may be a repulsive
gravitational force after the inclusion of
the back-reaction. However, this approach is
questionable since the validity of the
semiclassical theory is not well founded
for negative energy density cases. Here we are 
going to examine the validity of the correction to
the background metric calculated from
semiclassical theory for some range
of the physical parameters.
In order to get a quantitative measure of the deviation
from the semiclassical gravity theory, we here study 
the gravitational radiation by quantum systems in
\cite{Ford82}. We compare the predictions from
the full quantum theory and the semiclassical
theory in a linearized theory of gravity. We will
examine the coherent states, which can be
considered ``the most classical" quantum states.
This will give us some guidance in defining a
numerical measure of the deviation since we know
in the case of coherent states, this numerical 
measure should predict no violation of the
semiclassical theory.

We are stimulated by the discussion in \cite{Ford82} to 
propose a numerical measure for the applicability 
(or non-applicability) of semiclassical gravity theory
in various circumstances. We take the absolute 
value of the difference
$\langle\colon T_{\alpha\beta}(x_1)\,
T_{\mu\nu}(x_2)\colon\rangle-
\langle\colon T_{\alpha\beta}(x_1)\colon\rangle\,
\langle\colon T_{\mu\nu}(x_2)\colon\rangle$ 
first and then divide it by 
$\langle T_{\alpha\beta}(x_1)\,
T_{\mu\nu}(x_2)\rangle$ 
to form a dimensionless
quantity. The reason we choose the denominator 
to be $\langle\colon T_{\alpha\beta}(x_1)\,
T_{\mu\nu}(x_2)\colon\rangle$ 
and not 
$\langle\colon T_{\alpha\beta}(x_1)\colon\rangle\,
\langle\colon T_{\mu\nu}(x_2)\colon\rangle$
is to avoid artificial blowup when
$\langle\colon T_{\alpha\beta}(x)\colon\rangle$ 
vanishes for some physical range of parameters.

We propose that the extent to which the semiclassical approximation 
is violated can be measured by
the dimensionless quantity \cite{Kuo93,Kuo94}
\begin{equation}
\Delta_{\alpha\beta\mu\nu}(x,y)
\equiv \Biggl| 
{\langle\colon
T_{\alpha\beta}(x)\,T_{\mu\nu}(y)\colon\rangle
-\langle\colon T_{\alpha \beta}(x)\colon\rangle
\langle\colon T_{\mu \nu}(y)\colon\rangle 
\over \langle\colon T_{\alpha\beta}(x)\,
T_{\mu\nu}(y)\colon\rangle }  \Biggr| .
\end{equation}
This quantity is a dimensionless measure of the stress tensor fluctuations.
(Note that it is not a tensor, but rather the ratio of tensor components.) 
If its components are always small compared to unity, then these fluctuations
are small and we expect the semiclassical theory to hold. 
However, the numerous components and the dependence upon two spacetime points
make this a rather cumbersome object to study. For simplicity, we will
concentrate upon the coincidence limit, $x \rightarrow y$, of the purely 
temporal component of the above quantity, that is
\begin{equation}
\Delta(x)	\equiv \Biggl| 
{\langle\colon T_{00}{}^2(x)\colon
\rangle-\langle\colon T_{00}(x)\colon\rangle^2 
\over
\langle\colon T_{00}{}^2(x)\colon\rangle} 
\Biggr|.    
\label{eq:Delta} 
\end{equation}
The local energy density fluctuations are small when $\Delta \ll1$,
which we take to be a measure of the validity of the semiclassical theory.
Note that we have used normal ordering with respect to the Minkowski vacuum 
state to define the various operators.
 
To derive the form of the metric of the 
spacetime around an idealized straight string, 
we start with the Nambu action for an infinitely 
thin relativistic line \cite{Scherk75}
\begin{equation}
S=-\mu\int dA=-\mu\int d^2\sigma{\sqrt{-\gamma}},
\end{equation}
where $\mu$ is the linear density of the string
and 
\begin{equation}
\gamma_{ab}=\partial_a x^\mu(\sigma)\,
\partial_b x^\nu(\sigma)\,g_{\mu\nu}(x(\sigma))
\end{equation}
is the metric on the world sheet of the string
embedded in the background spacetime with metric
$g_{\mu\nu}(x(\sigma))$. Normally the magnitude 
of the linear density or equally the tension 
depends on the energy scale of the symmetry 
breaking which is responsible for the creation
of the string. For a symmetry breaking at the
grand unification scale, we have $\mu\approx 
(10^{16}\,GeV)^2\approx 10^{22}\,{\rm g\,cm}^{-1}$. 
The stress tensor of the string can be readily 
obtained by variation of the Nambu action with respect
to the metric,
\begin{equation}
T_{\mu\nu}(x)=-\,{2\over{\sqrt{-g}}}
{{\delta S}\over{\delta g^{\mu\nu}}}
\Big|_{g^{\mu\nu}=\eta^{\mu\nu}}.
\end{equation}
We can choose the coordinates $(\tau,\sigma)$
on the world sheet of the cosmic string such
that
\begin{eqnarray}
{\dot x}\cdot x'=0 \\
{\dot x}^2+x'^2=0,
\end{eqnarray}
where $x^i$ are the trajectories of the string.
In the Lorentz-Hilbert gauge, the linearized
Einstein equations 
can be solved. The $00$ component of $h_{\mu\nu}$
($g_{\mu\nu}=\eta_{\mu\nu}+h_{\mu\nu}$) is just twice 
the classical Newtonian potential. For a straight
string on the $z$-axis, the only non-zero components 
of the stress tensor are $T_{00}=-T_{33}=
\mu\,\delta^2({\bf x})$. From this, it is readily 
obtained that the classical Newtonian gravitational
potential vanishes in this spacetime 
\cite{Vilenkin87}. Assuming 
some cutoff distance (presumably the symmetry 
breaking scale when the string is created) of the 
Nambu action and
taking terms up to second order in $G_N\mu$, 
the metric of the spacetime around an idealized 
infinite straight string is solved to be
\begin{equation}
ds^2=-dt^2+dr^2+(1-4\,G_N\mu)^2\,
r^2d\varphi^2+dz^2.
\label{eq:StringMetric}
\end{equation}
If we make the substitution
$\theta=|1-4\,G_N\mu|\,\varphi$, the metric can be
recast into Minkowskian form. But the
periodicity in the angular coordinate is
changed into $\alpha\equiv 2\pi|1-4\,G_N\mu|$. 
In short, it is a conical space with deficit 
angle $8\pi G_N\mu$.

\section{Conformal Scalar Field in a Conical 
Space}

In this section we are going to talk about the
quantum field theory of a massless scalar field
in the conical space around a long straight
cosmic string, and the back-reaction of the 
scalar field to the conical space due to 
vacuum polarization. The stress tensor we use
here will be the so-called ``new improved 
energy-momentum tensor" in order to compare 
with other works. We will see that the change 
of the form of the stress tensor will not change 
our former conclusion concerning the relation
between the negativeness of the energy density
and the violation of the semiclassical theory
drastically.   
In the calculation, we use Feynman's propagator 
instead of the Hadamard elementary function to 
illustrate the independence of the choice of
the Green's function in obtaining a local form
of the stress tensor and the fluctuations.
Usually the form of the stress tensor can be 
reduced to a form containing only a single undetermined
function of the linear mean density by using
axial symmetry, Lorentz symmetry along the 
$z$-axis, conformal scale invariance, tracelessness
and conservation of energy-momentum
\cite{DeWitt89,Frolov87,Frolov89,Hiscock87}. 
However, we do not have 
this luxury here since the quantum fluctuations 
will not respect the original symmetry.

The dynamical equation of a massless scalar 
field in this metric is 
\begin{equation}
\Box\phi\equiv
\left[-{\partial^2\over \partial t^2}+
{1\over r}{\partial\over \partial r}
\left( r{\partial\over\partial r}\right) +
{1\over r^2}
{\partial^2\over \partial \theta^2}+
{\partial^2\over \partial z^2}\right]\,\phi=0
\end{equation}
The Feynman Green's function obeys the
equation 
\begin{equation}
\left[-{\partial^2\over \partial t^2}+
{1\over r}{\partial\over \partial r}
\left(r{\partial\over \partial r}\right) +
{1\over r^2}{\partial^2\over \partial \theta^2}+
{\partial^2\over \partial z^2}\right]\,G_F(x,
x') = -g^{-1/2}\,\delta(x,x') 
\end{equation}
In the following text we will omit the
subscript F in $G_F$ for simplicity.
The renormalized ``new improved stress 
tensor'' \cite{Callan70,Chernikov68} of 
the scalar field may be found from $G(x,x')$
through
\begin{equation}
\langle T_{\mu\nu}\rangle_{Ren}=
-i\lim_{x'\rightarrow x}
\left({2\over 3}\nabla_\mu\nabla_{\nu'}-{1\over 3}
\nabla_\mu\nabla_\nu-{1\over 6}g_{\mu\nu}
\nabla_\rho \nabla^{\rho'}\right)\,G_{Ren}(x,x'),
\end{equation}
where $G_{Ren}(x,x')$ is the renormalized 
Green's function, which will be determined 
later.
The Green's function may be obtained by
Schwinger's formalism, 
\begin{equation}
G(x,x')=i\int_0^\infty ds\,
e^{is\Box}\,\delta(x,x').
\end{equation}
We then expand the $\delta$ function as a 
sum and integral of the mode functions,
which is the completeness condition,
\begin{equation}
\delta(x,x')=
{i\over\alpha}\int_{-\infty}^\infty 
{d\omega\over 2\pi}\int_{-\infty}^\infty dp
\sum_{n=-\infty}^\infty u(x,\omega,k,p,n)\,
u^\ast(x',\omega,k,p,n).
\end{equation}
Here 
\begin{equation}
u(x,\omega,k,p,n)=(p/\alpha)^{1/2}\,
J_{|2n\pi/\alpha|}(pr)\,e^{i(kz-\omega t)}\,
e^{i(2n\pi\theta/\alpha)}
\end{equation}
are the eigenfunctions of the differential 
operator
\begin{equation}
\Box\equiv
-{\partial^2\over \partial t^2}+
{1\over r}{\partial\over \partial r}
(r{\partial\over \partial r}) +{1\over r^2}
{\partial^2\over \partial \theta^2}+
{\partial^2\over \partial z^2}, 
\end{equation}
and $J$ is 
the Bessel function of the first kind, $n$ is
an integer, $\omega$ and $k$ are arbitrary
real numbers, and $p$ is real and positive.
Replace this into the Schwinger's proper time 
representation of Green's function,
\begin{eqnarray}
G(x,x')&=
&{i\over\alpha}\int_{-\infty}^\infty 
{d\omega\over 2\pi}\int_{-\infty}^\infty p\,dp
\int_{-\infty}^\infty ds\,
e^{is(\omega^2-k^2-p^2)}\,
e^{-i\omega (t-t')}\, e^{ik(z-z')}\times 
\nonumber \\
&&\sum_{n=-\infty}^\infty 
J_{|2n\pi/\alpha|}(pr)\, J_{|2n\pi/\alpha|}(pr')
\, e^{i(2n\pi/\alpha)(\theta-\theta')}.
\label{eq:StringGreenFcn}
\end{eqnarray}
In contrast, the Minkowski Green's function is
\begin{eqnarray}
G_0(x,x')&=&{i\over{4\pi^2\,(x-x')^2}}\nonumber\\
&=&{i\over{4\pi^2\,\left[-(t-t')^2+r^2+r'^2+
(z-z')^2+2rr'\cos(t-t')\right]}}.
\end{eqnarray}
Since the processes of taking coincidence limits 
in the $t$, $r$ and $z$ directions commute 
with the operation of the stress tensor operator, 
we have the essential quantity
\begin{equation}
G(\theta,\theta')\equiv
\lim_{(t',r',z')\rightarrow (t,r,z)}
G(x,x')={i\over 4\alpha^2 r^2}\,
csc^2\left({\pi(\theta-\theta')\over\alpha}\right).
\end{equation}
Taking $\alpha\rightarrow 2\pi$ we have the 
Minkowskian correspondence (beware of a mistake
in the coefficient in \cite{Helliwell86})
\begin{equation}
G_0(\theta,\theta')\equiv
\lim_{(t',r',z')\rightarrow (t,r,z)}
G_0(x,x')
={i\over 16 \pi^2 r^2}
csc^2\left({(\theta-\theta')\over 2}\right).
\end{equation}
Thus the renormalized Green's function should be
\begin{equation}
G_{Ren}(\theta,\theta')=
G(\theta,\theta')-G_0(\theta,\theta').
\end{equation}

Due to the form of (\ref{eq:StringGreenFcn}) and 
the dependence of the Green's function on
the geodesic distance, all the derivatives needed
for calculating the stress tensor can be 
obtained from the derivative with respect to
$\theta$. They are
\begin{eqnarray}
\lim_{x'\rightarrow x}\partial_t^2 G(x,x')&=&
-\lim_{x'\rightarrow x}\partial_t\partial_{t'}G(x,x')=
\lim_{x'\rightarrow x}\partial_z\partial_{z'}G(x,x')=
-\lim_{x'\rightarrow x}\partial_z^2 G(x,x')\nonumber\\
&=&{1\over 3r^2}\,
\lim_{\theta'\rightarrow \theta}
(1+\partial_{\theta}^2) \,
G(\theta,\theta'),
\end{eqnarray}
and
\begin{eqnarray}
\lim_{x'\rightarrow x}\partial_r\partial_{r'}G(x,x')
={1\over 3r^2}\,
\lim_{\theta'\rightarrow \theta}
\left(4+\partial_{\theta}^2\right)\, 
G(\theta,\theta').
\end{eqnarray}
Using
\begin{equation}
\lim_{\theta'\rightarrow \theta}
{\partial^2 \over \partial{\theta}^2}\, 
G_{Ren}(\theta,\theta')={i\over 480\pi^2\,r^4}\,
\left[ \left(1-4G_N\mu\right)^{-4}-1\right],
\end{equation}
we can get the renormalized energy density
\begin{eqnarray}
\langle\colon T_{00}\colon\rangle &=&
-i\lim_{x'\to x}\left({2\over 3}
\partial_t\partial_{t'}
-{1\over 3}\partial_t^2+{1\over 6}g_{00}
\nabla_\rho\nabla^{\rho'}
\right) G_{Ren}(x,x')\nonumber\\
&=&-i\lim_{\theta'\to\theta}\left[ 
-{1\over 3r^2}\left( 1+{\partial^2\over
\partial\theta^2}\right)\right] G_{Ren}(\theta,\theta')
\nonumber\\
&=&-\,{1\over 1440\pi^2\,r^4}\,
\left[ \left(1-4G_N\mu\right)^{-4}-1\right] .
\end{eqnarray}
The stress tensor can be shown to be
\begin{equation}
\langle\colon T_{\mu\nu}\colon\rangle
={1\over 1440\pi^2\,r^4}\,
\left[ \left(1-4G_N\mu\right)^{-4}-1\right]
\,{\rm diag}[-1,1,-3r^2,1].
\end{equation}
It is worthwhile to notice that due to the 
boundary effect, the energy density is
negative for physical regions (i.e., where the
Nambu action holds) of the
parameter $\mu$. Reinserting $\hbar$ and $c$, 
we have
\begin{equation}
\rho=\langle\colon T_{00}\colon\rangle=
-\,{\hbar \over 1440\pi^2\,r^4\,c}\,
\left[ \left(1-{4G_N\mu\over c^2}\right)^{-4}-1\right]
\approx -10^{-4}{G_N\mu\hbar\over c^3\, r^4}.
\end{equation}
We know that the linear mean density of the 
cosmic string produced at the grand unified scale
gives $G_N\mu\approx 10^{-6}$.
The numerical value is $\rho\approx -{10^{-47}
{\rm cm}\,{\rm g} \over r^4}$.
And the assumption of the Nambu action should be
valid up to the symmetry breaking scale.
Assuming $r$ can be of the order of the grand 
unification scale $L\approx 10^{-30}$ cm, the upper 
limit of the value of the energy density is about 
$-10^{73} {\rm g}\,{\rm cm}^{-3}$, a pretty 
high density. Previously we have shown 
that for quantum states with negative energy 
density, the semiclassical gravity theory 
may not be trusted. The amount of deviation from
the semiclassical theory for a scalar field in the 
vicinity of a cosmic string should be examined
carefully to see if the semiclassical theory can 
be applied.

The expectation value of the square of
energy density becomes \cite{Kuo93,Kuo94}
\begin{eqnarray}
\langle T_{00}{}^2(x)\rangle_{Ren}=
&&\lim_{x_1,x_2,x_3,x_4\to x}\,
\left({2\over 3}\partial_{t_1}\partial_{t_2}-
{1\over 3}\partial_{t_1}\partial_{t_1}
-{1\over 6}g_{\mu\nu}
\nabla_{\rho_1}\nabla^{\rho_2}\right)
\nonumber \\
&&\left({2\over 3}\partial_{t_3}\partial_{t_4}-
{1\over 3}\partial_{t_3}\partial_{t_3}-
{1\over 6}g_{\mu\nu}
\nabla_{\sigma_3}\nabla^{\sigma_4}\right)
\nonumber\\
&&\Bigl[\,G_{Ren}(x_1,x_2)\,G_{Ren}(x_3,x_4)+
G_{Ren}(x_1,x_3)\,G_{Ren}(x_2,x_4)+\nonumber\\
&&G_{Ren}(x_1,x_4)\,G_{Ren}(x_2,x_3)\,\Bigr] .
\end{eqnarray}
The result is 
\begin{eqnarray}
\langle T_{00}{}^2(x)\rangle_{Ren}&=&
\langle T_{00}(x)\rangle^2_{Ren}+
{{{{G_N^2\mu}^2}\,{{\left(1-2\,G_N\mu\right)}^2}\,
 }\over 
{97200\,{{\left(1-4\,G_N\mu \right) }^8}\,{{\pi}^4}\,{r^8}}}
\nonumber\\
&&\times\left(-2541 + 35748\,G_N\mu-199432\,{{G_N^2\mu}^2}  
\right. \nonumber\\
&&\left.
+511744\,{{G_N^3\mu}^3}-511744\,{{G_N^4\mu}^4}\right) .
\end{eqnarray}
The second term in the equation is the fluctuating 
part. Due to the smallness of the value of the 
factor $G_N\mu$, the second term can be approximated
by terms proportional to $G_N^2\mu^2$, of the same
order of $\langle T_{00}(x)\rangle^2_{Ren}$.
So even though it is inversely proportional to the eighth 
power of the distance from the string, it is not
negligible in the semiclassical description.
From the formula for $\Delta$, we have 
\begin{eqnarray}
\Delta={ {2541-35748 G_N\mu +199432
\left( G_N\mu \right)^2 - 
511744\left(G_N\mu\right)^3
+511744\left( G_N\mu\right)^4 }\over
{2529-35652G_N\mu+199048\left(G_N\mu\right)^2- 
510976\left(G_N\mu\right)^3+
510976\left( G_N\mu \right)^4 } }\, \nonumber\\
\end{eqnarray}
which is independent of $r$.
The degree of the negativeness of the energy
density depends both on the parameter $G_N\mu$
in the theory and the distance from the string.
However, the measure $\Delta$ does not depend on
the distance or the symmetry breaking scale $\mu$
crucially far from Planck scale, which is similar 
to the calculation for the two conducting plates, where 
$\Delta$ is independent of the separation of the 
plates. Besides, since $G_N \mu$ is pretty small,
$\Delta\approx 1$. That is, no matter how far
away from the string, semiclassical theory should
not be trusted. Even though the quantum fluctuations get
smaller with increasing distance, they
are still comparable to the energy density.

From the dependence of $\Delta$ to $G_N\mu$ and $r$
in Planck units($G_N\mu\rightarrow \mu$), we can 
easily see the value of $\Delta$ is small and 
insensitive to both $r$ and $\mu$ when away from
Planck scale. The measure of deviation $\Delta$ 
is large when the value of the energy density
is negative, independent of $r$. That means the 
deviation from semiclassical gravity theory is 
significant at all distances from the string.

\section{CONCLUSION AND DISCUSSION}

From this investigation we know that the back-reaction
in conical spacetime can not be calculated from 
the naive semiclassical gravity theory.

Using a criterion obtained before, we are able to
tell the range of applicability quantitatively. It is
interesting to see that the degree of violation of the
semiclassical theory does not depend on the linear
density of the cosmic string (or equivalently, the
symmetry breaking scale) or the distance from the string 
critically. This result indicate the back-reaction 
calculation in \cite{Hiscock87} based on the semiclassical
theory of gravity should not to be trusted.

This is a good example in which semiclassical gravity 
theory can be violated even far from Planck scale.
Negative energy density may be a sign for such violation.
Once we encounter negative energy density, the validity
of the semiclassical theory should be checked and 
some other extensions describing the fluctuating 
nature of the stress tensor and hence the spacetime
should be used.   

The interesting possible indications of the present 
result in the structure formation mechanism is to 
be investigated.

\section*{Acknowledgement}
The author would like to thank Professors L. H. Ford, 
T. Roman, A. Vilenkin, and E. Everett for helpful 
discussions.

\vfill 

\end{document}